\RequirePackage{ifpdf}
\ifpdf 
\documentclass[pdftex]{sigma}
\else
\documentclass{sigma}
\fi

\begin{document}
\allowdisplaybreaks

\renewcommand{\PaperNumber}{031}

\FirstPageHeading

\ShortArticleName{$q$-Deformed Bi-Local Fields II}

\ArticleName{$\boldsymbol{q}$-Deformed Bi-Local Fields II}

\Author{Haruki TOYODA~$^\dag$ and Shigefumi NAKA~$^\ddag$}
\AuthorNameForHeading{H. Toyoda and S. Naka}

\Address{$^\dag$~Laboratory of Physics, College of Science and Technology Nihon University,\\
$\phantom{^\dag}$~7-24-1 Narashinodai Funabashi-shi Chiba, Japan}

\EmailD{\href{mailto:toyoda@phys.ge.cst.nihon-u.ac.jp}{toyoda@phys.ge.cst.nihon-u.ac.jp}}

\Address{$^\ddag$~Department of Physics, College of Science and
Technology
Nihon University,\\
 $\phantom{^\ddag}$~1-8-14 Kanda-Surugadai Chiyoda-ku Tokyo, Japan}

\EmailD{\href{mailto:naka@phys.cst.nihon-u.ac.jp}{naka@phys.cst.nihon-u.ac.jp}}

\ArticleDates{Received December 01, 2005, in f\/inal form February
22, 2006; Published online March 02, 2006}

\Abstract{We study a way of $q$-deformation of the bi-local
system, the two particle system bounded by a relativistic harmonic
oscillator type of potential, from both points of view of mass
spectra and the behavior of scattering amplitudes. In our
formulation, the deformation is done so that $P^2$, the square of
center of mass momentum, enters into the deformation parameters of
relative coordinates. As a result, the wave equation of the
bi-local system becomes nonlinear with respect to $P^2$; then, the
propagator of the bi-local system suf\/fers signif\/icant change
 so as to get a convergent self energy to the second order.
 The study is also made on the covariant $q$-deformation in four dimensional spacetime.}


\Classification{32G07; 81R50; 81R60}

\section{Introduction}

The aim of studying non-local f\/ield theories proposed by
Yukawa~\cite{non-local} was originally twofold: f\/irstly, to
derive characteristic properties of elementary particles such as
the mass spectrum of hadrons, from their extended structure;
secondly, to deal with the divergence dif\/f\/iculty, which is
inherent in local f\/ield theories with local interactions. The
bi-local f\/ield theory~\cite{bi-local} was the f\/irst attempt by
Yukawa following this line of thought. For such f\/ields, the
f\/irst aim has been addressed in many works in the context of
ef\/fective relativistic two-particle systems of quark and
anti-quark bound systems. In particular, two-particle systems
bounded by a relativistic harmonic oscillator potential were
useful in deriving the linear mass-square spectrum associated with
the Regge behavior in their scattering
amplitude~\cite{Barger-Cline, Regge}. Contrastingly, there has
been little success in the pursuit of the second aim, mainly
because the bi-local f\/ields are reduced to a~superposition of an
inf\/inite number of local f\/ields with dif\/ferent masses,
although some people have claimed that the second order
self-energy becomes convergent associated with the direction of
the center-of-mass momentum. In addition, the problems of the
unitarity of the scattering matrix and causality are also serious
for those f\/ields, because a bi-local system in general allows
time-like relative motion. Usually, such a degree of freedom is
frozen by an additional subsidiary condition~\cite{Takabayasi}.
However, this is not always successful
 for interacting cases. This situation may be dif\/ferent from that of
 string models that are characterized by the Virasoro condition
 associated with the parameterization invariance in such an extended model;
 nevertheless the study of bi-local f\/ield theories does not come to end,
 since a small change in models, sometimes, will cause a signif\/icant change in their physical properties.

Given the situation described above, the purpose of this paper is
to study the $q$-deformation~\cite{Macfarlane, Wess, Sogami,
Quantum-Groups}
 of a bi-local system characterized by a relativistic harmonic oscillator potential,
  because the $q$-deformation is well-def\/ined for harmonic oscillator systems.
  In a previous paper~\cite{5-dimension}, we studied a $q$-deformed 5-dimensional spacetime
  such that the extra dimension generates a~harmonic oscillator type of
  potential for particles embedded in that spacetime. Then, the propagator
  of the particles in that spacetime acquires a signif\/icant convergent property
  by requiring that the 4-dimensional spacetime variables and the extra dimensional
  spacetime variable are mixed by the deformation. It then happens that the
4-dimensional spacetime variables do not commute with the f\/ifth
space variables. We can expect that the same situation arises in
the bi-local system, if we carry out the deformation of the
relative variables in this system along the line of the
$q$-deformed 5-dimensional spacetime. This extension of the
bi-local systems was done in~\cite{q-bi-local}; and, we could show
that the bi-local system was free from the divergence in one-loop
level. In this stage, however, we could not f\/ind the way of
$q$-deformation being consistent with the Lorentz covariance. In
this paper, thus, we intend adding a new sight for the covariant
deformation, in addition to the review of the $q$-deformed
bi-local f\/ields.

In the next section, we formulate the bi-local system.\ In
Section~3 we construct the $q$-de\-formed bi-local system with the
$q$-deformed relative coordinates. In this case, we def\/ine the
deformation so as to get a Lorentz invariant resultant wave
equation. In Section~4, the interaction of the bi-local f\/ield is
discussed in the context of calculating Feynman diagrams. Some
scattering amplitudes between the bi-local system and external
scalar f\/ields are also studied by considering their Regge
behavior. A second order self-energy diagram is also calculated to
study the convergence of the model. Section~5 is devoted to
summary and discussion. In Appendix~A, several ways of
$q$-deformation in $N$-dimensional oscillator variables are
discussed to f\/ind a possible way to the covariant deformation.
In Appendix~B, we add a new attempt to construct a~phase space
action of the $q$-deformed bi-local system. Appendix~C is devoted
to giving of the outline for calculating of a one-loop diagram
corresponding to the self-energy of the $q$-deformed bi-local
f\/ield.

\section[Bi-local field theories]{Bi-local f\/ield theories}

The bi-local theory obtains a potential approach to the bound
state of relativistic two particle system. Then the classical
action of an equal mass two-particle system that leads to the
standard bi-local f\/ield equations is given by~\cite{bi-local}
\begin{gather*}
S= \int d\tau\frac{1}{2} \sum_{i=1}^{2}
\left\{ \frac{\dot{x}^{(i)2}}{e_i}+e_i (m^2+V(\bar{x})) \right\},
\end{gather*}
where $\bar{x}=x^{(1)}-x^{(2)}$ is the relative coordinators of
the system. Here $e_i$'s are einbein that guarantee the
invariance of $S$ under the reparametrization of $\tau$. By taking
the variation of $S$ with respect to $e_i$, we obtain the
equal-mass constraints
\begin{gather*}
\frac{\delta S}{\delta e_i}=-\big\{p^{(i)2}-(m^2 +
V(\bar{x}))\big\}=0, \qquad i=1,2,
\end{gather*}
where $p^{(i)}_\mu=\frac{\delta S}{\delta x^{(i)\mu}}$ are the
momenta conjugate to $x^{(i)\mu}$. These constraints can be
rewritten as
\begin{gather}
\frac{1}{2}P^2+2\big(\bar{p}^2 -m^2 \big)-2V(\bar{x})=0,\label{constrain2} \\
P\cdot p=0, \label{constraint1}
\end{gather}
where $P=p^{(1)}+p^{(2)}$ and
$\bar{p}=\frac{1}{2}\big(p^{(1)}-p^{(2)}\big)$ are the total
momentum and the relative momentum of the bi-local system,
respectively. The variables $X_\mu$ satisfying the canonical
commutation relations $[P_\mu,X_\nu]=ig_{\mu\nu}$,
$[\bar{p}_\mu,X_\nu]=0$ are determined uniquely to be
$X_\mu=\frac{1}{2}\big(x^{(1)}_\mu+x^{(2)}_\mu\big)$, which are
nothing but the center of mass coordinates of the equal-mass
bi-local
 system\footnote{${\rm diag}\,(g_{\mu\nu})=(+---)$.}.

Now, one can verify that these constraints are compatible for the
covariant harmonic oscillator potential $V(\bar{x})=-k^2\bar{x}^2
$ in the following sense: In $q$-number theory, we can introduce
the oscillator variables def\/ined by
\begin{gather*}
\bar{x}=\sqrt{\frac{1}{2k}}\, (a^{\dag}+a),\qquad \bar{p}=i\sqrt{\frac{k}{2}}\, (a^{\dag}-a).
\end{gather*}
Then reading equation~(\ref{constraint1}) in the sense of
expectation $\langle \phi |P\cdot p|\phi\rangle=0$ as in the
Gupta--Bleuler formalism in QED, equations~(\ref{constrain2}) and
(\ref{constraint1}) can be understood respectively as the master
wave equation of the bi-local system and the physical state
condition; that is, we can put
\begin{gather}
\left( \alpha' P^2 +\frac{1}{2}\{a_{\mu}^\dag , a^{\mu}\}-\omega \right)|\phi\rangle= 0,\label{master}\\
P^\mu a_\mu |\phi\rangle =0,\label{subsidiary}
\end{gather}
where $\alpha'=\frac{1}{8\kappa}$ and $\omega =\frac{m^2}{2k}$. In
this stage, the compatibility between equations~(\ref{master}) and
(\ref{subsidiary}) becomes clear. After the second quantization of
physical states $\{|\phi\rangle \}$, those become the $q$-number
f\/ields, so called `the bi-local f\/ield'.

  We also note that subsidiary condition (\ref{constraint1}) def\/ines the bi-local
  f\/ield associated with the indef\/inite metric formalism of the Lorentz group.
  If we read equation~(\ref{constraint1}) as $P_\mu a^{\mu\dag}|\phi\rangle=0$ instead
  of~(\ref{subsidiary}), then we have a def\/inite metric formalism of bi-local f\/ield theories.
  The former has a~simi\-larity to the string models and the latter shows an interesting form
  factor as meson-like bound states~\cite{Takabayasi}.

\section[$q$-deformed bi-local system]{$\boldsymbol{q}$-deformed bi-local system}

The $q$-deformed one-dimensional harmonic oscillator is def\/ined
by the oscillators and the number operator satisfying
\begin{gather*}
a_qa_q^{\dag}-qa_q^{\dag}a_q= q^{-N}, \qquad [N,a_q^{\dag}]=a_q^{\dag},
\end{gather*}
The oscillator $(a_q,a_q^{\dag})$ can be realized from the
standard harmonic oscillator $(a,a^{\dag})$, the
$(a_q,a_q^{\dag})$ with $q=1$, through the mapping
\begin{gather}
a_q=a\sqrt{\frac{[N]_q}{N}},\qquad N=a^\dag a,\qquad
[N]_q=\frac{\sinh(N\log q)}{\sinh(\log q)},\label{mapping}
\end{gather}
There are several ways of $q$-deformation in four-dimensional
harmonic oscillators; and, the following is a similar way to
(\ref{mapping}) in constructing $q$-deformed oscillators:
\begin{gather*}
a_{q\mu} =a_{\mu}\sqrt{\frac{[N]_q}{N_\mu}},\qquad
[N]_q=\frac{\sinh\{(N+\beta+\frac{1}{2})\log q\}}{\sinh(\frac{1}{2}\log q)},\nonumber \\
N=-a^{\dag}_{\mu}a^{\mu}, \qquad N_0 =-a_0^{\dag}a^0, \qquad N_i=a_i^{\dag}a^i.
\end{gather*}
Here, the $\beta$ may be an operator commuting with
$(a_\mu,a_\mu^\dag)$; and so, it may contain the center of mass
momentum $P_\mu$. Interesting result is obtained by putting $\beta
=\beta_0 - \frac{3}{4}\alpha'P^2$ by considering the covariance of
a resultant wave equation. Then, the master wave equation becomes
\begin{gather}
\left(\alpha'P^2 +
\frac{1}{2}\{a_{q\mu},a_{q}^{\mu\dagger}\}(P^2,N)-\omega\right)|\Phi\rangle
. =0,\label{q-master}
\end{gather}
from which we can obtain the mass-square like operator
\begin{gather*}
m^2=-\frac{1}{2\alpha'}\left\{a_{q \mu},a^{\mu\dag}_q\right\}(P^2)
+\frac{\omega}{\alpha'}=-\frac{2}{\alpha'}\frac{\sinh[(N+\beta+\frac{1}{2})\log
q]} {\sinh[\frac{1}{2}\log q]}+\frac{\omega}{\alpha'}.
\end{gather*}
It should be noticed that the $m^2$ operator is a Lorentz scalar
in spite of that the mapping breaks the Lorentz covariance.
However, if we do not worry about the complexity of $[N]$, we can
adopt a covariant mapping, which yields the same master wave
equation as (\ref{q-master}). We will discuss separately this
problem in Appendix~A, since such a mapping is important from the
viewpoint of the Lorentz covariance.

\begin{figure}[h]
\centerline{ \includegraphics[width=5.5cm]{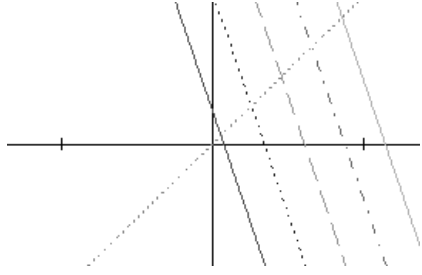}}
  \caption{The line $y=x$ vs $y=-\frac{1}{2\alpha'}\{a_q,a_q^\dag\}+\frac{\omega}{\alpha'}$
  with $n=0,1,2,\ldots$. The mass-square eigenvalues are represented by the $x$ coordinates of the intersection.}
\end{figure}

Now, the mass square eigenvalues $\{m^2_n\}$ are obtained by
solving the non-linear equation (\ref{q-master}) with respect to
$P^2$ for each eigenvalue $n$ $(=0,1,2,\ldots)$ of $N$; that is,
those are solutions of $m^2_n =-
\frac{1}{\alpha'}\{a_q,a_q^\dag\}(m^2_n) +\frac{\omega}{\alpha'}$.
It should be noted that the $q$-deformed bi-local f\/ield is free
from space-like solutions for $\alpha>0$ as can be seen from
Fig.~1.

The above discussion implies that the free bi-local f\/ield is
similar to local free f\/ields in the sense that its spacetime
development can be determined by the Cauchy data  without
contradicting causality. Also, the Feynman propagator
\begin{gather*}
G(P^2,N)=\left( P^2 +\frac{1}{2\alpha'}\{a_{q\mu},a_q^{\mu\dag}\}
-\frac{\omega}{\alpha'}+i\epsilon \right)^{-1}
\end{gather*}
decreases exponentially as $N$, $|P^2|\rightarrow \infty$. As
shown in the next section, this enables us to obtain a
f\/inite-vacuum-loop amplitude, in contrast to the usual local
f\/ield theories. This property was observed f\/irstly
in~\cite{5-dimension}.

\section[Interaction of the bi-local field]{Interaction of the bi-local f\/ield}

We here discuss, shortly, on the interaction caused by the three
vertex of bi-local f\/ields. The symmetric three vertex in Fig.~2
can be naturally def\/ined by~\cite{Goto, Goto-Naka}
\begin{gather}\!\!\!
\left.
\begin{array}{c}
\big[x_{\mu}^{(1)}(b)-x_{\mu}^{(2)}(a)\big] |V\rangle= 0 \\
{}\big[p_{\mu}^{(1)}(b)+p_{\mu}^{(2)}(a)\big] |V\rangle= 0
\end{array}
\right\}(a,b,c~{\rm cyclic}).\label{3-vertex-1}
\end{gather}

The vertex function $|V\rangle$ determined by these equations,
however, does not satisfy the physical state condition $P\cdot
a(k)|V\rangle=0$, $(k=a,b,c)$; in order to get a physical vertex,
thus, we have to put $|V_{\rm phys}\rangle
=\Lambda_a\Lambda_b\Lambda_c|V\rangle $ with projection operators
def\/ined by $P\cdot
a(k)\Lambda_k=0,\Lambda_k^2=\Lambda_k,(k=a,b.c)$.

Using the vertex function, we can calculate the second order
scattering amplitudes in Fig.~4~$\sim$ Fig.~7. In the vertices of
those amplitudes, it is suf\/f\/icient to put one of bi-local
f\/ields in Fig.~2, say the particle `b', in the ground state,
which can be identif\/ied with an external scalar f\/ield as in
Fig.~3. Then the vertex conditions (\ref{3-vertex-1}) become
\begin{gather*}
[x^{(1)}(a)-x^{(2)}(c)]|V\rangle =[x^{(2)}(a)-x^{(1)}(c)]|V\rangle=0,\nonumber\\
[p^{(1)}(a)+p^{(2)}(c)]|V\rangle=[p^{(2)}(a)+p^{(1)}(c)]|V\rangle=0.
\end{gather*}

\begin{figure}[h]
\centerline{\begin{minipage}[t]{7.5cm}\centering
 \includegraphics[width=4.2cm]{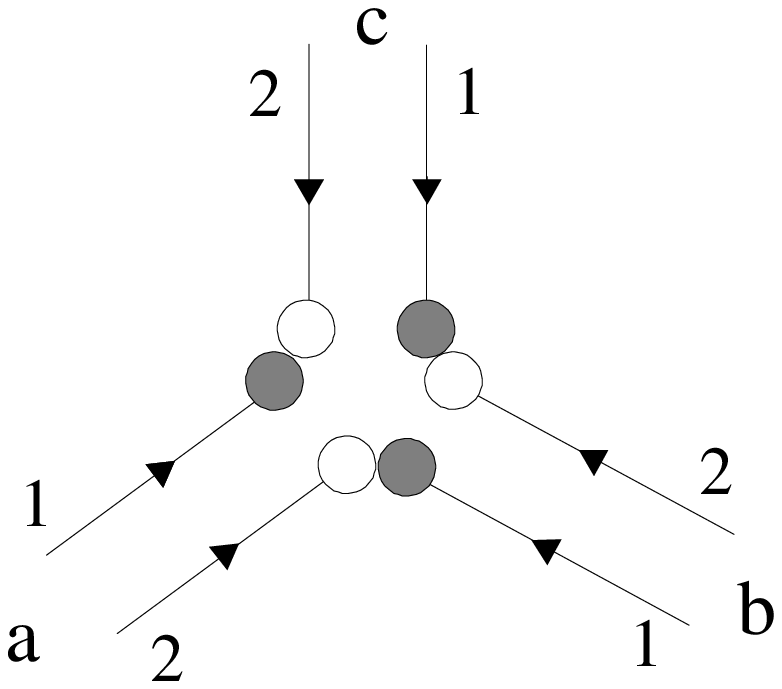}
 \vspace{-3mm}
  \caption{The a, b, c designate three bi-local systems; and 1, 2 are constituents in each bi-local system.}
\end{minipage}
\hfill
\begin{minipage}[t]{7.5cm}\centering
 \includegraphics[width=4.2cm]{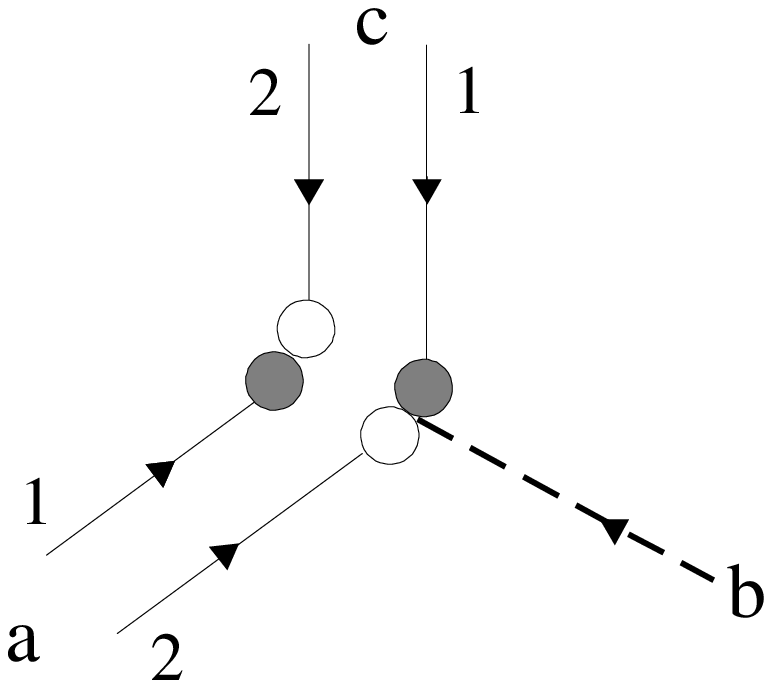}
 \vspace{-3mm}
  \caption{The dashed line denotes the local sca\-lar f\/ield corresponding to the ground state of~$b$.}
\end{minipage}}
\end{figure}

These equations with additional physical state conditions can be
solved easily to give
\begin{gather*}
|V\rangle=g \delta^{(4)}(P(a)+P(b)+P(c)) \nonumber \\
\phantom{|V\rangle=}{}\times \exp\left[
-\frac{i}{2}\sqrt{\frac{1}{2k}}(a(a)^\dag -a(c)^\dag)_\perp\cdot
P(b)+a(a)^\dag_\perp\cdot a(c)^\dag_\perp \right]|0\rangle .
\end{gather*}
where $a_{\perp\mu}=O_{\mu\nu}a^{\nu}$ with
$O_{\mu\nu}=\eta_{\mu\nu}-\frac{p_\mu\cdot p_\nu}{p^2}$. With aid
of this vertex function, one can calculate the second order
scattering amplitudes of bi-local f\/ields. In particular, for a
simpler case with ground states $a$, $a'$, it is found that the
$s$-channel scattering amplitude decreases rapidly for
$s\rightarrow \infty$ with $t$ f\/ixed and the $t$-channel
scattering amplitude shows the Regge behavior for $t \rightarrow
\infty$ with $s$ f\/ixed. This means that the $t$-channel
amplitude in Fig.~5 obtained by interchanging~$s$ and~$t$ from the
second-order scattering amplitude exhibits Regge behavior for
large $s$, with $t$ f\/ixed~\cite{q-bi-local}.

\begin{figure}[h]
\centerline{\begin{minipage}[b]{7.5cm}\centering
 \includegraphics[width=2.8cm]{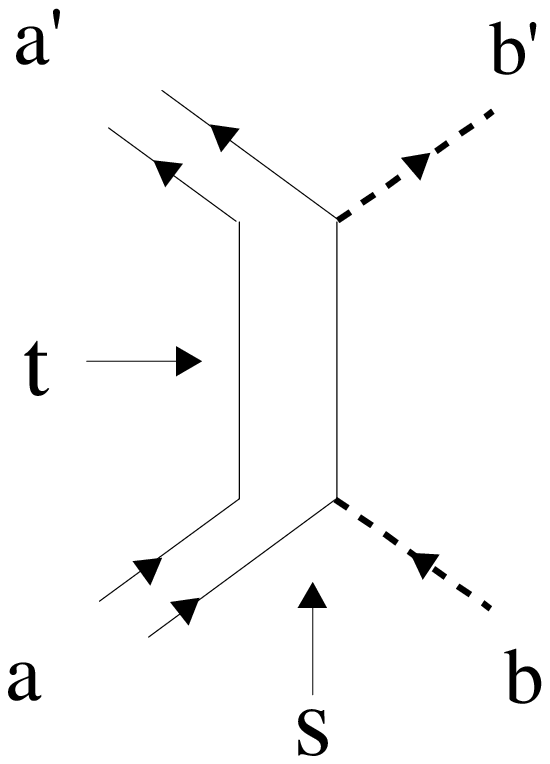}
 \vspace{-3mm}
  \caption{A second-order scat\-tering amplitude of a
   bi-local system by two external f\/ields with $s=(P(a)+P(b))^2$ and $t=(P(a)-P(a'))^2$.}
\end{minipage}
\hfill
\begin{minipage}[b]{7.5cm}\centering
 \includegraphics[width=4.3cm]{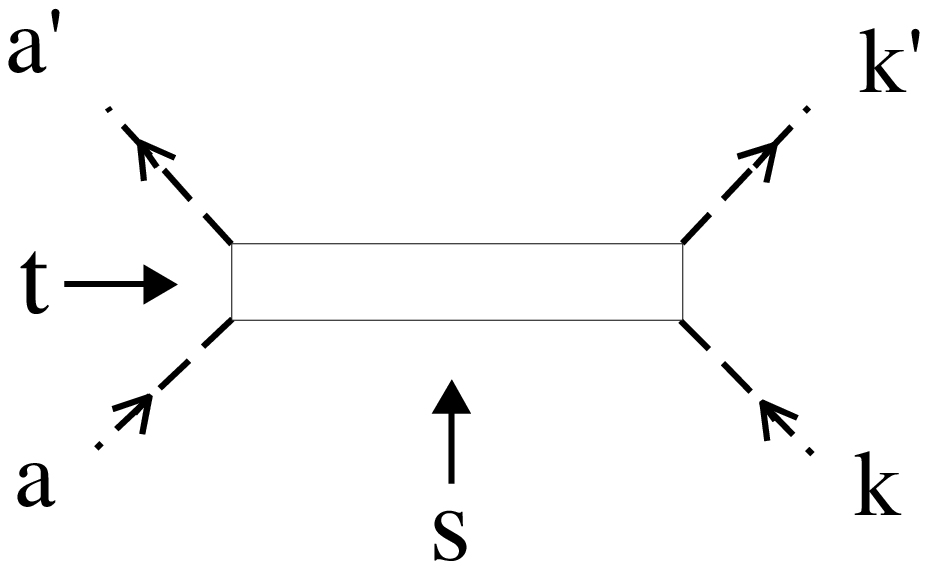}
 \vspace{-3mm}
  \caption{The $t$-channel amplitude obtained through interchange $s\leftrightarrow t$ from Fig.~4.\newline $\,$}
\end{minipage}}
\end{figure}

A more remarkable fact in this model is that the loop diagram in
Fig.~7 corresponding to the self energy of the ground state of the
bi-local system is found to be convergent \cite{q-bi-local}.
Indeed, according to the magnitude of momenta in the internal
line, the self energy can be written as
\begin{gather}
\delta m^2\sim \delta m^2(|p|\langle |k|)+\delta m^2(|p|\gg
|k|).\label{loop}
\end{gather}

\begin{figure}[h]
\centerline{\begin{minipage}[b]{7.5cm}\centering
 \includegraphics[width=2.8cm]{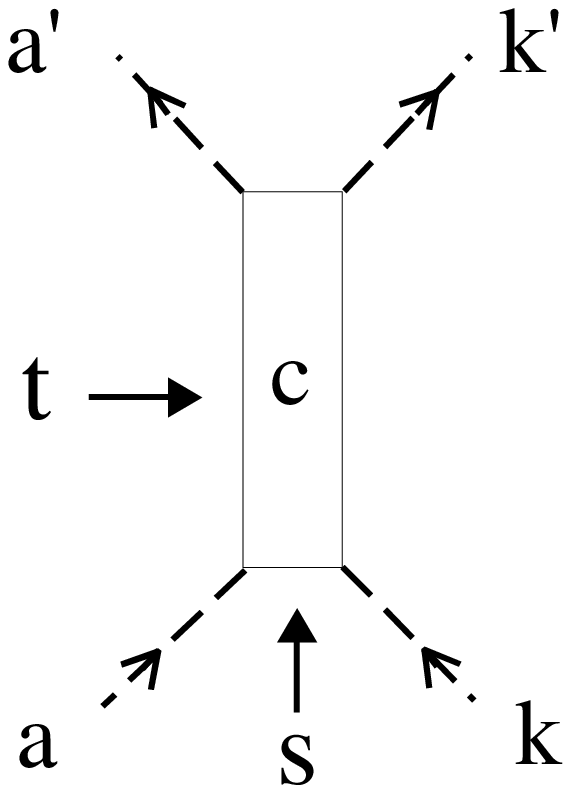}
 \vspace{-3mm}
  \caption{The $s$-channel scattering amplitude for four grand states.}
\end{minipage}
\qquad\quad
\begin{minipage}[b]{7.5cm}\centering
 \includegraphics[width=5cm]{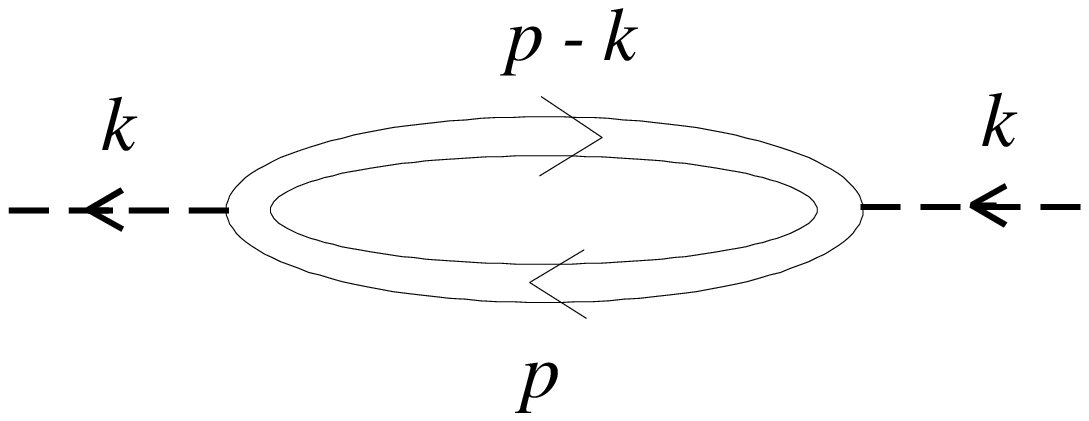}
 \vspace{-3mm}
  \caption{The self-energy for the grand state. \newline ~~~$\,$}
\end{minipage}}
\end{figure}

The f\/irst term of equations~(\ref{loop}) is f\/inite obviously,
and the second term can be evaluated as
\begin{gather*}
\delta m^2(|p|\gg|k|)\sim i\left(
\frac{g\pi\alpha\sinh(\frac{1}{2}\log q)} {2\alpha'_0 \log
q}e^{(\beta_0 +\frac{1}{2})\log q} \right)^2,
\end{gather*}
Therefore, the self-energy in our model is found to be convergent
by virtue of
 the factor $\sinh (\cdot)$ in the propagator with the deformation parameter~$q\neq 1$.

\section{Summary and discussion}

The relativistic two-particle system, the bi-local system, bounded
by a 4-dimensional harmonic oscillator potential yields a
successful description of two-body meson-like states. The
mass-square spectrum, then, arises from the excitation of relative
variables, which are independent of the center-of-mass variables.

In this work, we have constructed a $q$-deformed bi-local system
with harmonic oscillator type bound potential. The deformation
$(a_{q\mu}(P^2),a_{q\mu}^\dag(P^2))$ is carried out so as to
include the center-of mass momentum in the deformation parameters.
Then $q$-deformed relative coordinates are non-commutative with
each other, while the center-of-mass coordinates remain as
commutative variables. The formal mass square operator is Lorentz
invariant, though the mapping $(a_\mu,a_\mu^\dag)\rightarrow
(a_{q\mu},a_{q\mu}^\dag)$ spoils the covariance. However, the way
of mapping giving the same mass square operator is not unique; and
there is a covariant mapping in compensation of simple commutation
relations of the oscillator variables.

In the $q$-deformed bi-local system, the mass eigenvalues are real
simple zeros of master wave equation; then, the propagator becomes
free from multi-pole ghosts. Further, the wave function of system
acquires new properties such that  the propagator of system damped
rapidly as $|P^2|$ tends to inf\/inity. We have also studied the
interaction of the bi-local system with external scalar f\/ield,
which are identif\/ied with the ground state of the system. Then,
we could verify the following: First, the second-order $t$-channel
scattering amplitude exhibits Regge behavior in the limit
$t\rightarrow \infty$ with $s$ f\/ixed. Secondly, a second-order
loop diagram, which corresponds to a self-energy of the system,
shows convergent property providing $q\neq 1$. This is due to
a~characteristic property of the propagator.

To complete the $q$-deformed bi-local theory within the framework
of f\/ield theories, it is neces\-sary to study the higher-order
diagrams in addition to the analysis on the causality. Those are
interesting future problems.

\newpage

\appendix

\section[Representation of a $q$-oscillator]{Representation of a $\boldsymbol{q}$-oscillator}

We here discuss the representation of the $q$-oscillator variables
def\/ined by
\begin{gather}
 [A,A^\dagger]_q \equiv A A^\dagger -q^\alpha A^\dagger A =q^{-\alpha ( N + \beta) }, \label{q-commutator}
\end{gather}
where $\alpha$, $\beta$ and $q$ $(\geq 1)$ are real parameters,
and $N=a^\dagger a$ is the number operator for the ordinary
oscillator variables satisfying $[a,a^\dagger]=1$. There is a
mapping between $(a,a^\dagger)$ and $(A,A^\dagger)$ such as
\begin{gather*}
 A = a \sqrt{\frac{[N]_q}{N}},\qquad A^\dagger = \sqrt{\frac{[N]_q}{N}}a^\dagger ,
\end{gather*}
where $N=a^\dagger a$. To determine $[N]_q$, let us recall
$Na=a(N-1)$ and $Na^\dagger=a^\dagger(N+1)$; then, we can verify
\begin{gather*}
 A A^\dagger = [N+1]_q,\qquad A^\dagger A = [N]_q ,
\end{gather*}
from which equation~ (\ref{q-commutator}) can be reduced to
\begin{gather*}
 [N+1]_q - q^\alpha [N]_q = q^{-\alpha (N +\beta) } . 
\end{gather*}
This recurrence equation can be solved easily; and we have
\begin{gather}
 [N]_q = \frac{q^{\alpha(N+\beta)}-q^{-\alpha(N+\beta)}}{q^\alpha -q^{-\alpha}}. \label{[N]}
\end{gather}
with the condition $[0]=0$.

Under this $q$-deformation, the Hamiltonian for ordinary
oscillator $\frac{1}{2}\{ a^\dagger ,a \}+\beta$ will be replaced
by $\frac{1}{2} \{ A^\dagger ,A \}$, which can be rewritten as
\begin{gather*}
 \frac{1}{2} \{ A^\dagger,A \} = \frac{1}{2}\left( [N]_q + [N+1]_q \right)
  = \frac{1}{2}\frac{\sinh \left[\alpha(N + \beta +\frac{1}{2})\log q \right]}
  {\sinh(\frac{1}{2}\alpha\log q)}. 
\end{gather*}
Indeed, one can verify that $\frac{1}{2} \{ A,A^\dagger \}
\rightarrow \frac{1}{2}\{ a^\dagger ,a \}+\beta$ according as as
$q \rightarrow 1$. This implies that $\beta$ plays the role of an
additional term in the zero-point energy; and, the $\alpha$ can be
absorbed into the def\/inition of $q$ by the substitution
$q^\alpha \rightarrow q$.

The $q$-deformation can be extended to $D$-dimensional oscillator
variables def\/ined by $[a_i,a_j^\dagger]=\delta_{ij}$,
$[a_i,a_j]=[a_i^\dagger,a_j^\dagger]=0$ $(i,j=1,2,\ldots,D)$. We
note that there is no mapping $A_i(a,a^\dagger)$ satisfying
\begin{gather*}
 [A_i,A_j^\dagger]_q=\delta_{ij}f(N)\qquad {\rm and}\qquad [A_i,A_j]_q=[A_i^\dagger,A_j^\dagger]_q=0,
\end{gather*}
where $N=\sum_i a_i^\dagger a_i$. Indeed,
 since $A_j[A_i,A_j^\dagger]_q=q^{2\alpha}[A_i,A_j^\dagger]_q A_j+q^\alpha [A_i,f(N)]$
 for $i\neq j$, we have $[A_i,f(N)]=0$, which holds only when $A_i$ is a
 function of $N$, that is, a function without vector indices $\{i\}$.
 The mappings listed below, however, may be useful for the model building.

Case~(i): $A_i=a_i\sqrt{\frac{[N_i]_q}{N_i}}$. The \{$A_i\}$
satisfy the simple algebra
\begin{gather*}
 [A_i,A_j^\dagger]_q =\delta_{ij}q^{-\alpha ( N_i + \beta) }. 
\end{gather*}
However, the mapping does not preserve $U(D)$ symmetry, even for
$\sum_i\{A_i,A_i^\dagger\}$.

Case~(ii):  $A_i=a_i\sqrt{\frac{[N]_q}{N}}$. This mapping
preserves the $U(D)$ vector property of $A_i$, while the algebra
of $\{A_i\}$ is modif\/ied so that
\begin{gather*}
 A_iA_j^\dagger-\left(\frac{[N+1]_q}{[N]_q}\frac{N}{N+1}\right)A_j^\dagger A_i =\frac{[N+1]_q}{N+1}\delta_{ij}.
\end{gather*}
Further, with $[N]$ in equation~(\ref{[N]}), the Hamiltonian
becomes a summed form:
\begin{gather*}
 \frac{1}{2}\sum_i\{ A_i^\dagger,A_i \} = \frac{1}{2}
 \frac{\sinh \left[\alpha(N + \beta +\frac{1}{2})\log q \right]}
 {\sinh(\frac{1}{2}\alpha\log q)}+\frac{1}{2}\frac{D+1}{N+1}
 \frac{\sinh \left[\alpha(N + \beta +1)\log q \right]}{\sinh(\alpha\log q)}.
\end{gather*}
We shall bring up this case later, again, since this case is
interesting in the viewpoint of the covariant formulation of a
$q$-deformed bi-local model.

Case~(iii): $A_i = a_i\sqrt{\frac{[N]_q}{N_i}}$. We can then
verify that the $\{A_i\}$ satisfy a $U(D)$ covariant algebra such
as
\begin{gather*}
 [A_i,A_j^\dagger]_q = q^{-\alpha(N+\beta)}A_i[N]_q^{-1}A_j^\dagger, 
\end{gather*}
although the mapping spoils that symmetry; and the Hamiltonian
becomes an invariant form, which is used in Section~3:
\begin{gather}
\frac{1}{2}\sum_i\{ A_i^\dagger,A_i \} =\frac{D}{2}\frac{\sinh
\left[\alpha(N + \beta +\frac{1}{2})\log q \right]}
{\sinh(\frac{1}{2}\alpha\log q)}. \label{q-hamiltonian-2}
\end{gather}

We now turn to the covariant mapping (ii) to note that there is a
function $[N]$ related with a given Hamiltonian $H(N)$ in
principle. In order to show this, we f\/irst write
$\frac{1}{2}\sum_i\{A_i^\dag,A_i\}=H(N)$ as
\begin{gather*}
 [N+1]+\frac{N+1}{N+D}[N]=\frac{2(N+1)}{N+D}H(N),
\end{gather*}
in which $N$ may be read as an integer. Then using the function
$\varphi(N)$ def\/ined by
\begin{gather*}
 \varphi(N)=\prod_{k=0}^N\frac{k+1}{k+D}=(D-1)B(D-1,N+2),
\end{gather*}
with the convention $\varphi(0)=1$, we can get the recurrence
relation
\begin{gather*}
 \frac{[N+1]}{\varphi(N)}+\frac{[N]}{\varphi(N-1)}=2\frac{H(N)}{\varphi(N-1)}\equiv g(N).
\end{gather*}
This equation can be solved by the standard manner; and we obtain
\begin{gather*}
 [N]=(-1)^{N-1}\varphi(N-1)\left\{ [1]+\sum_{k=1}^{N-1}(-1)^kg(k)  \right\}
\end{gather*}
The right-hand side of this equation is determined by $H(N)$ only.
Therefore, if it is necessary, we can arrive at the expression
(\ref{q-hamiltonian-2}) starting from the covariant mapping (ii),
though the $[N]$ becomes complex one.

\section[The phase-space action for a $q$-deformed bi-local model]{The phase-space action for
a $\boldsymbol{q}$-deformed bi-local model}

It is well known that the phase-space action for the
one-dimensional harmonic oscillator has a~simple form
\begin{gather}
  L(z,\dot{z}) = \frac{i}{2}z^* \stackrel{\leftrightarrow}{\partial_t} z-\langle z|H|z\rangle,
  \qquad H=\frac{\hbar\omega}{2}\{a^\dag,a\},\label{p-action-1}
\end{gather}
where $|z\rangle=e^{-|z|^2/2+za^\dag|0\rangle}$ is coherent with
$\langle z|z\rangle=1$. Indeed, if we substitute
$z=\sqrt{\frac{m\omega}{2}}x +\frac{i}{\sqrt{2m\omega}}p$ for
(\ref{p-action-1}), then the $L$ will be reduced to the standard
action of the harmonic oscillator after eliminating~$p$. In a
similar sense, for the $q=1$ bi-local model, the action can be
written as
\begin{gather*}
 L = -P\cdot\dot{X} +\frac{i}{2}z_\mu^* \stackrel{\leftrightarrow}{\partial_\tau} z^\mu  -eH(P,z,z^*) ,
\end{gather*}
where
\begin{gather}
 H(P,z,z^*) = \alpha^\prime P^2+\langle\bar{z}|\frac{1}{2}\{a_\mu^\dag,a^\mu\}|z\rangle-\omega, \label{H-1}
\end{gather}
and $e(\tau)$ is the einbein, which guarantees the $\tau$
reparametrization invariance of the action. The total momentum
$P_\mu$ can be eliminated from $L$ obviously by using the
constraint $\frac{\partial}{\partial P^\mu}L=0$. The $H$ in the
$q$-deformed case is simply obtained by the substitution
$\frac{1}{2}\{a_\mu^\dag,a^\mu\} \rightarrow
\frac{1}{2}\{A_\mu^\dag,A^\mu \}$ in equation~(\ref{H-1}).

Let us show that even in this $q$-deformed case, we can eliminate
$P_\mu$ from $L$ by using the constraint
\begin{gather*}
\frac{\partial L}{\partial P^\mu} = -\dot{X_\mu}+e\partial_{P^\mu}
H =0.
\end{gather*}
If we notice here that $(\partial_{P} H)^2\equiv f(P^2)$ is a
function of $P^2$ only, the c onstraint allow us to solve~$P^2$ in
terms of $(e^{-1}\dot{X})^2$; then, we have
\begin{gather*}
  P^2 = f^{-1}\left[ \left(e^{-1}\dot{X}\right)^2 \right]\qquad
  \mbox{and}\qquad P\cdot\dot{X}=  \sqrt{f^{-1}\left[\left(e^{-1}\dot{X}\right)^2\right]\dot{X}^2}.
\end{gather*}
Substituting this expression for $L$, we f\/inally obtain
\begin{gather*}
 L=-\sqrt{f^{-1}\left[\left(e^{-1}\dot{X}\right)^2\right]\dot{X}^2}+
+\frac{i}{2}z_\mu^* \stackrel{\leftrightarrow}{\partial_\tau}
z^\mu +
eH\left(f^{-1}\left[\left(e^{-1}\dot{X}\right)^2\right],z,\dot{z}
\right).
\end{gather*}
One can verify that the constraint corresponding to the master
wave equation is obtained by taking the derivative of~$L$ with
respect to~$e$.

\section{Loop amplitude}

We here show the outline calculation of  the loop diagram
corresponding to the self-energy $\delta m^2$ for the ground
state, which can be written as
\begin{gather}
 \delta m^2 \sim g^2\int d^4p \,{\rm Tr}\left[G((p-k)^2,N_\perp(p-k)):
 e^{-\frac{i}{2}\bar{x}\cdot O(p-k)\cdot k}:G(p^2,N_\perp(p)):
 e^{\frac{i}{2}\bar{x}\cdot O(p)\cdot k}: \right] \nonumber \\
 \phantom{\delta m^2}{}= \delta m^2(|p|\lnsim |k|) + \delta m^2(|p|\gg |k|) . \label{self-energy}
\end{gather}
The f\/irst term of the r.h.s.\ in equation~(\ref{self-energy})
will be f\/inite, and second term can be roughly evaluated as
\begin{gather}
\delta m^2(|p| \gg |k|) \sim g^2\int d^4p \int_C
\frac{d\zeta}{2\pi i\zeta}
\int_C \frac{d\zeta'}{2\pi i\zeta'} \sum_{n=0}^\infty \sum_{n'=0}^\infty  \nonumber \\
\qquad {}\times {\rm Tr}_{\rm
phys}\left[\zeta^{N_\perp(p)}:e^{-\frac{i}{2} \bar{x}\cdot
O(p)\cdot k}:\zeta'^{N_\perp(p)}:e^{\frac{i}{2}\bar{x}\cdot
O(p)\cdot k}: \right] \zeta^{-n}G(p^2,n)\zeta'^{-n'}G(p^2,n') ,
\label{self-energy-2}
\end{gather}
where ${\rm Tr}_{\rm phys}$ means the `trace' in the physical
subspace. The trace can be calculated by using the coherent state
$|z\rangle =e^{-z\cdot a^\dag}|0\rangle$ as follows
\begin{gather}
{\rm Tr}_{\rm phys}[\cdot] =
\int\left(\prod_{\mu=0}^3\frac{d^2z^\mu}{\pi}\right)
  e^{\bar{z}^*\cdot z}\langle \bar{z}|\cdots|z\rangle |_{z_\parallel=0} \nonumber \\
\phantom{{\rm Tr}_{\rm phys}[\cdot]}{} =
\frac{1}{(1-\zeta\zeta')^4}\exp \left[
-\frac{m_0^2}{2p^2}\frac{2\zeta\zeta'-(\zeta +
\zeta')}{1-\zeta\zeta'}\right] \simeq \frac{1}{(1-\zeta\zeta')^4},
\label{trace}
\end{gather}
where $\bar{z}=(-z^0,z^1,z^2,z^3)$ and $z_\parallel=\frac{P(P\cdot
z)}{P^2}$. The last form of equation~(\ref{trace}) seems to be
singular at $\zeta=\zeta'^{-1}$ $(\zeta'=\zeta^{-1})$, which is,
however, located in outside of the counter $C$. Further, since
$G(p^2,n)$ has no poles of imaginary $p^0$ \cite{q-bi-local}, we
can evaluate the integral with respect to $p$ in
equation~(\ref{self-energy-2}) by means of analytic continuation
$p=(i\bar{p}^0,\bar{p}^i)$.

Then, approximating the lower bound of the integration with
respect to $\bar{p}$ to $0$, we can carried out the $\bar{p}$
integration in equation~(\ref{self-energy-2}) explicitly; and, we
obtain
\begin{gather*}
 \delta m^2(|p| \gg |k|)  \sim i \left(\frac{g\pi\alpha'\sinh(\frac{1}{2}\log q)}
 {2\alpha_0'\log q}e^{-(\beta_0+\frac{1}{2})\log q}\right)^2. 
\end{gather*}

\subsection*{Acknowledgements}

The authors with to express their thanks to the members of their
laboratory for discussions and encouragement.

\LastPageEnding

\end{document}